\documentstyle[12pt]{article}
\catcode`@=12
\begin{document}
\thispagestyle{empty}
\begin{flushright} UCRHEP-T183\\December 1996\
\end{flushright}
\vspace{0.5in}
\begin{center}
{\Large \bf Broken Supersymmetric U(1) Gauge Factor\\ at the TeV Scale\\}
\vspace{1.2in}
{\bf Ernest Ma\\}
\vspace{0.3in}
{\sl Department of Physics, University of California\\
Riverside, CA 92521, USA\\}
\vspace{1.2in}
\end{center}
\begin{abstract}\
The appearance of a broken supersymmetric U(1) gauge 
factor at the TeV scale is relevant for several reasons.  If it truly exists, 
then one important consequence is that at the 100 GeV energy scale, the 
two-doublet Higgs structure is of a more general form than that of the 
Minimal Supersymmetric Standard Model (MSSM).  This is a prime example 
of tree-level nondecoupling.  Furthermore, a particular $U(1)_N$ from the 
superstring-inspired $E_6$ model allows for the existence of naturally 
light singlet neutrinos which may be necessary to accommodate the totality 
of neutrino-oscillation experiments.
\end{abstract}
\vspace{0.3in}
\noindent {\it To appear in Proc. of the Third International Workshop on 
Particle Physics Phenomenology, Chin Shan, Taiwan (November, 1996)}
\newpage
\baselineskip 24pt
\section{Why Supersymmetric U(1)?}

Consider the top-down approach.  Start with the $E_8 \times E_8$ heterotic 
string, which compactifies to the $E_6$ superstring,\cite {1} which is then 
broken by flux loops transforming as the adjoint {\bf 78} representation. 
Under $SU(3)_C \times SU(3)_L \times SU(3)_R$,
\begin{equation}
{\bf 78} = (3, 3, 3^*) + (3^*, 3^*, 3) + (8, 1, 1) + (1, 8, 1) + (1, 1, 8).
\end{equation}
Now (1,8,1) can be used for $SU(3)_L \rightarrow SU(2)_L \times U(1)_{Y_L}$ 
and (1,1,8) for $SU(3)_R \rightarrow U(1)_{T_{3R} + Y_R}$ so that the 
correct $Q = T_{3L} + Y_L + T_{3R} + Y_R$ is obtained, hence $U(1)_{Y_L} 
\times U(1)_{T_{3R} + Y_R}$ remains unbroken and can be rewritten as 
$U(1)_Y \times U(1)_\eta$.  The extra $U(1)_\eta$ presumably survives\cite{2} 
down to the TeV energy scale where it is broken together with the 
supersymmetry.

Consider also the bottom-up approach.  Start with a possible experimental 
clue, such as the $R_b \equiv (Z \rightarrow b \bar b / Z \rightarrow 
hadrons)$ excess.  Look for an U(1) explanation, and remarkably $U(1)_\eta$ 
is also found.\cite{3}  Another possible clue is the totality of 
neutrino-oscillation experiments (solar, atmospheric, and laboratory) 
which suggest that there are at least 4 neutrinos.  Details of how this 
is related to an extra U(1) will be presented later.

\section{Tree-Level Nondecoupling at the 100 GeV Scale}

As the U(1) gauge factor is broken together with the supersymmetry at the 
TeV scale, the resulting heavy scalar particles have nondecoupling 
contributions to the interactions of the light scalar particles.\cite{4}
Consequently, the two-doublet Higgs structure is of a more general form 
than that of the MSSM.  Previous specific examples have been given.\cite{5}
Here I present the most general analysis.\cite{6} Let
\begin{eqnarray}
\tilde \Phi_1 &=& \left( \begin{array} {c} \bar \phi_1^0 \\ - \phi_1^- 
\end{array} \right) \sim (1,2, -{1 \over 2}; -a), \\  \Phi_2 &=& \left( 
\begin{array} {c} \phi_2^+ \\ \phi_2^0 \end{array} \right) \sim (1,2, 
{1 \over 2}; -1 + a), \\ \chi &=& \chi^0 \sim (1,1,0;1),
\end{eqnarray}
where each last entry is the arbitrary assignment of that scalar multiplet 
under the extra U(1), assuming of course that the superpotential has 
the term $f \Phi_1^\dagger \Phi_2 \chi$.  The corresponding scalar 
potential contains thus
\begin{equation}
V_F = f^2 [(\Phi_1^\dagger \Phi_2)(\Phi_2^\dagger \Phi_1) + (\Phi_1^\dagger 
\Phi_1 + \Phi_2^\dagger \Phi_2)\bar \chi \chi],
\end{equation}
and from the gauge interactions,
\begin{eqnarray}
V_D &=& {1 \over 8} g_2^2 [(\Phi_1^\dagger \Phi_1)^2 + (\Phi_2^\dagger 
\Phi_2)^2 + 2 (\Phi_1^\dagger \Phi_1)(\Phi_2^\dagger \Phi_2) - 
4 (\Phi_1^\dagger \Phi_2) (\Phi_2^\dagger \Phi_1)] \nonumber \\ 
&+& {1 \over 8} g_1^2 [\Phi_1^\dagger \Phi_1 - \Phi_2^\dagger \Phi_2]^2 
\nonumber \\ &+& {1 \over 2} g_x^2 [-a \Phi_1^\dagger \Phi_1 - (1-a) 
\Phi_2^\dagger \Phi_2 + \bar \chi \chi]^2.
\end{eqnarray}
Let $\langle \chi \rangle = u$, then $\sqrt 2 Re \chi$ is a physical scalar 
boson with $m^2 = 2 g_x^2 u^2$, and the $(\Phi_1^\dagger \Phi_1) \sqrt 2 Re 
\chi$ coupling is $\sqrt 2 u (f^2 - g_x^2 a)$. Hence the effective 
$(\Phi_1^\dagger \Phi_1)^2$ coupling $\lambda_1$ is given by
\begin{eqnarray}
\lambda_1 &=& {1 \over 4} (g_1^2 + g_2^2) + g_x^2 a^2 - {{2(f^2-g_x^2 a)^2} 
\over {2 g_x^2}} \nonumber \\ &=& {1 \over 4} (g_1^2 + g_2^2) + 2 a f^2 - 
{f^4 \over g_x^2}.
\end{eqnarray}
Similarly,
\begin{eqnarray}
\lambda_2 &=& {1 \over 4} (g_1^2 + g_2^2) + 2 (1-a) f^2 - {f^4 \over g_x^2}, 
\\ \lambda_3 &=& -{1 \over 4} g_1^2 + {1 \over 4} g_2^2 + f^2 - {f^4 \over 
g_x^2}. \\ \lambda_4 &=& -{1 \over 2} g_2^2 + f^2,
\end{eqnarray}
where the two-doublet Higgs potential has the generic form
\begin{eqnarray}
V &=& m_1^2 \Phi_1^\dagger \Phi_1 + m_2^2 \Phi_2^\dagger \Phi_2 + m_{12}^2 
(\Phi_1^\dagger \Phi_2 + \Phi_2^\dagger \Phi_1) + {1 \over 2} 
\lambda_1 (\Phi_1^\dagger \Phi_1)^2 \nonumber \\ &~& 
+ ~{1 \over 2} \lambda_2 (\Phi_2^\dagger 
\Phi_2)^2 + \lambda_3 (\Phi_1^\dagger \Phi_1)(\Phi_2^\dagger \Phi_2) + 
\lambda_4 (\Phi_1^\dagger \Phi_2)(\Phi_2^\dagger \Phi_1).
\end{eqnarray}
From Eqs.~(7) to (10), it is clear that the MSSM is recovered in the limit 
$f = 0$.  Let $\langle \phi_{1,2}^0 \rangle \equiv v_{1,2}$, $\tan \beta 
\equiv v_2/v_1$, and $v^2 \equiv v_1^2 + v_2^2$, then this $V$ has an upper 
bound on the lighter of the two neutral scalar bosons given by
\begin{equation}
(m_h^2)_{max} = 2 v^2 [\lambda_1 \cos^4 \beta + \lambda_2 \sin^4 \beta 
+ 2 (\lambda_3 + \lambda_4) \sin^2 \beta \cos^2 \beta] + \epsilon,
\end{equation}
where we have added the radiative correction due to the $t$ quark and its 
supersymmetric scalar partners, {\it i.e.}
\begin{equation}
\epsilon = {{3 g_2^2 m_t^4} \over {8 \pi^2 M_W^2}} \ln \left( 1 + {\tilde m^2 
\over m_t^2} \right).
\end{equation}
Using Eqs.~(7) to (10), we obtain
\begin{equation}
(m_h^2)_{max} = M_Z^2 \cos^2 2 \beta + \epsilon + {f^2 \over 
{\sqrt 2 G_F}} \left[ A - {f^2 \over g_x^2} \right],
\end{equation}
where
\begin{equation}
A = {3 \over 2} + (2 a - 1) \cos 2 \beta - {1 \over 2} \cos^2 2 \beta.
\end{equation}
If $A > 0$, then the MSSM bound can be exceeded.  However, $f^2$ is still 
constrained from the requirement that $V$ be bounded from below.  For a given 
$g_x$, we can vary $a$, $\cos 2 \beta$, and $f$ to find the largest 
numerical value of $m_h$, which grows with $g_x$.  In a typical model 
such as the $U(1)_\eta$ model, $g_x^2 = (25/36) g_1^2 \simeq 0.09$ and 
$a = 1/5$, for which $m_h < 142$ GeV.  If we increase $g_x^2$ to 0.5, 
and allow all other parameters to vary, then we get $m_h < 190$ GeV.

\section{Framework for a Naturally Light Singlet Neutrino}

There are at present a number of neutrino experiments with data\cite{7,8,9} 
which can be interpreted as being due to neutrino oscillations.  Solar 
data\cite{7} indicate the oscillation of neutrinos differing in the square 
of their masses of the order $\Delta m^2 \sim 10^{-5} ~{\rm eV}^2$ for the 
matter-enhanced solution or $\Delta m^2 \sim 10^{-10} ~{\rm eV}^2$ for the 
vacuum solution.  Atmospheric data\cite{8} indicate possible oscillation of 
$\Delta m^2 \sim 10^{-2} ~{\rm eV}^2$.  More recently, the liquid scintillator 
neutrino detector (LSND) experiment has obtained results\cite{9} which 
indicate possible oscillation of $\Delta m^2 \sim 1 ~{\rm eV}^2$.  

To accommodate all the above data as being due to neutrino oscillations, 
it is clear that four neutrinos are needed to have three unequal mass 
differences.  Since the invisible width of the $Z$ boson is already 
saturated with the three known doublet neutrinos $\nu_e$, $\nu_\mu$, and 
$\nu_\tau$, {\it i.e.} from $Z \rightarrow \nu \bar \nu$, one must then 
have a fourth neutrino which does not couple to the $Z$ boson, {\it i.e.} 
a singlet.  The question is why such a singlet neutrino should be light.

The lightness of doublet neutrinos is canonized by the seesaw mechanism:
\begin{equation}
{\cal M}_{\nu,N} = \left[ \begin{array} {c@{\quad}c} 0 & m_D \\ m_D & m_N 
\end{array} \right] \Rightarrow m_\nu \sim {m_D^2 \over m_N}.
\end{equation}
To have a light singlet $S$, assume an extra U(1) gauge factor as well as 
2 doublets $(\nu_E, E)_L)$ and $(E^c, N_E^c)_L$ such that
\begin{equation}
{\cal M}_{\nu_E, N_E^c, S} = \left[ \begin{array} {c@{\quad}c@{\quad}c} 
0 & m_E & m_1 \\ m_E & 0 & m_2 \\ m_1 & m_2 & 0 \end{array} \right] 
\Rightarrow m_S \sim {{2 m_1 m_2} \over m_E}.
\end{equation}
It is thus desirable to have an extra U(1) under which $N$ is trivial but 
$S$ is not.  The two sectors must also be connected so that oscillations 
may occur between $\nu$ and $S$.  If we now assume the Higgs scalars of this 
theory to carry the quantum numbers of $(\nu_E,E)$, $(E^c, N_E^c)$, and $S$, 
then the combined mass matrix is given by
\begin{equation}
{\cal M} = \left[ \begin{array} {c@{\quad}c@{\quad}c@{\quad}c@{\quad}c} 
0 & m_D & 0 & m_3 & 0 \\ m_D & m_N & 0 & 0 & 0 \\ 0 & 0 & 0 & m_E & m_1 \\ 
m_3 & 0 & m_E & 0 & m_2 \\ 0 & 0 & m_1 & m_2 & 0 \end{array} \right],
\end{equation}
which reduces to 
\begin{equation}
{\cal M}_{\nu,S} = \left[ \begin{array} {c@{\quad}c} m_D^2/m_N & m_1 m_3/m_E 
\\ m_1 m_3/m_E & 2 m_1 m_2/m_E \end{array} \right]
\end{equation}
as desired.  Other considerations such as anomaly cancellation and simplicity 
implies\cite{10} that we have the supersymmetric $SU(3)_C \times SU(2)_L 
\times U(1)_Y \times U(1)_N$ model\cite{11} with particle content given by 
the fundamental {\bf 27} representation of $E_6$.

\section{The U(1)-Extended Supersymmetric Model}

The conventional decomposition of $E_6$ is as follows.  First we have 
$E_6 \rightarrow SO(10) \times U(1)_\psi$, with
\begin{eqnarray}
(16,1) &=& (u,d) + u^c + e^c + d^c + (\nu_e,e) + N, \\ (10,-2) &=& h + 
(E^c,N_E^c) + h^c + (\nu_E,E), \\ (1,4) &=& S,
\end{eqnarray}
where $2 \sqrt 6 Q_\psi$ has been denoted.  Next we have $SO(10) 
\rightarrow SU(5) \times U(1)_\chi$.  In general, a linear combination of 
$U(1)_\psi$ and $U(1)_\chi$ may survive down to the TeV energy scale.  Let 
$Q_\alpha = \cos \alpha Q_\psi - \sin \alpha Q_\chi$, then for 
$\tan \alpha = -1/\sqrt {15}$, we have $Q_\alpha (N) = 0$ which is what 
we want.  Call this $U(1)_N$, then
\begin{equation}
2 \sqrt {10} Q_N = 6 Y_L + T_{3R} - 9 Y_R.
\end{equation}
Specifically, under $U(1)_N$, we have
\begin{equation}
(u,d), u^c, e^c \sim 1; ~~~ d^c, (\nu_e,e) \sim 2; ~~~ N \sim 0;
\end{equation}
\begin{equation}
h, (E^c, N_E^c) \sim -2; ~~~ h^c, (\nu_E, E) \sim -3; ~~~ S \sim 5.
\end{equation}
Comparing against Eq.~(2) to (4), we find $a = 3/5$ and $g_x^2 = (25/24) 
g_1^2$.  The largest numerical value of $m_h$ in this case is 140 GeV.

Assume 3 copies of the {\bf 27} representation to accommodate the 3 families 
of quarks and leptons, and impose a discrete $Z_2$ symmetry such that one 
copy of $(\nu_E,E)$, $(E^c,N_E^c)$, and $S$ superfields are even, and all 
others are odd.\cite{11}  The scalar components of the $Z_2$-even 
superfields develop nonzero vacuum expectation values: $\langle \tilde S 
\rangle = u$, $\langle \tilde \nu_E \rangle = v_1$, and $\langle \tilde N_E^c 
\rangle = v_2$.  Neglecting $v_{1,2}$ for the time being, we find the mass of 
$Z'$ to be equal to that of the physical scalar boson $\sqrt 2 Re \tilde S$, 
{\it i.e.} $(\sqrt 5/2) g_N u$, where $g_N^2 = (5/3) g_1^2$ is a very good 
approximation obtained from normalizing $U(1)_Y$ and $U(1)_N$ at the 
grand-unification scale.\cite{12}  Furthermore, the corresponding $S$ pairs 
up with the $Z'$-gaugino to form the mass matrix
\begin{equation}
{\cal M}_{\tilde Z', S} = \left[ \begin{array} {c@{\quad}c} M_1 & M_{Z'} \\ 
M_{Z'} & 0 \end{array} \right],
\end{equation}
where $M_1$ is the soft supersymmetry breaking U(1) gaugino mass.  The other 
2 $S$'s are naturally light singlet neutrinos.  The $\cal M$ of Eq.~(18) is 
actually $12 \times 12$ because it contains 3 $\nu$'s and 2 $S$'s.  Because 
$Z'$ couples to $S$ according to Eq.~(25), the invisible width of $Z'$ is 
very much enhanced in this model.
\begin{equation}
{{\Gamma (Z' \rightarrow \nu \bar \nu)} \over {\Gamma (Z' \rightarrow 
\ell^- \ell^+)}} = {4 \over 5}; ~~~ {{\Gamma (Z' \rightarrow S \bar S)} 
\over {\Gamma (Z' \rightarrow \ell^- \ell^+)}} = {10 \over 3}.
\end{equation}
In all previous phenomenological studies of $Z'$ from $E_6$, the possibility 
of light $S$'s has not been recognized.  The above would serve as a 
distinctive signature of the $U(1)_N$ model.

\section{Z - Z' Sector}

The new $Z'$ of this model mixes with the standard $Z$.
\begin{equation}
{\cal M}^2_{Z, Z'} = \left[ \begin{array} {c@{\quad}c} g_Z^2 (v_1^2 + v_2^2)/2 
& g_N g_Z (-3v_1^2 + 2 v_2^2)/2 \sqrt {10} \\ g_N g_Z (-3v_1^2 + 2v_2^2)/2 
\sqrt {10} & g_N^2 (25u^2 + 9v_1^2 + 4v_2^2)/20 \end{array} \right].
\end{equation}
This results in a slight shift of the physical $Z$ mass and a slight change 
in its couplings to the usual quarks and leptons.  These deviations can be 
formulated in terms of the oblique parameters.\cite{13}
\begin{eqnarray}
\epsilon_1 &=& \left( \sin^4 \beta - {9 \over 25} \right) {v^2 \over u^2} 
~\simeq~ \alpha T, \\ \epsilon_2 &=& \left( \sin^2 \beta - {3 \over 5} \right) 
{v^2 \over u^2} ~\simeq~ - {{\alpha U} \over {4 \sin^2 \theta_W}}, \\ 
\epsilon_3 &=& {2 \over 5} \left( 1 + {1 \over {4 \sin^ \theta_W}} \right) 
\left( \sin^2 \beta - {3 \over 5} \right) {v^2 \over u^2} ~\simeq~ 
{{\alpha S} \over {4 \sin^2 \theta_W}},
\end{eqnarray}
where $v^2 \equiv v_1^2 + v_2^2$ and $\tan \beta \equiv v_2/v_1$.  Note that 
for $\sin^2 \beta$ near 3/5, $\epsilon_{1,2,3}$ are all suppressed.  In any 
case, the experimental errors on these quantities are fractions of a percent, 
hence $u \sim$ TeV is allowed.

\section{Conclusions}

(1) There are theoretical and phenomenological hints for the existence of an 
extra supersymmetric U(1) gauge factor which is broken at the TeV scale. 
(2) In particular, the $U(1)_N$ model from the $E_6$ superstring is very 
desirable for understanding the totality of neutrino-oscillation 
experiments.  (3) Because of tree-level nondecoupling in the scalar sector, 
the effect of an extra supersymmetric U(1) gauge factor is already felt by 
the two-doublet Higgs structure at around 100 GeV, which will be of a more 
general form than that of the Minimal Supersymmetric Standard Model.

\section*{Acknowledgments} I thank Chao-Qiang Geng and all the other local 
organizers for their great hospitality and a stimulating workshop. 
This work was supported in part by the U. S. Department of Energy under Grant 
No. DE-FG03-94ER40837.

\end{document}